\documentclass[aps,preprint]{revtex4}%
\usepackage{amssymb}
\usepackage{amsfonts}
\usepackage{amsmath}
\usepackage{graphicx}%
\providecommand{\U}[1]{\protect\rule{.1in}{.1in}}

\begin{document}
\title{Dimensional Analysis and the Time Required to Urinate}
\author{Peter Palffy-Muhoray}
\author{Yijing Chen}
\author{Hiroshi Yokoyama}
\affiliation{Liquid Crystal Institute, Kent State University}
\author{Xiaoyu Zheng}
\affiliation{Dept. of Mathematical Sciences, Kent State University}

\begin{abstract}
According to the recently discovered 'Law of Urination', mammals, ranging in
size from mice to elephants, take, on the average, 21s to urinate. We attempt
to gain insights into the physical processes responsible for this uniformity
using simple dimensional analysis. We assume that the biological apparatus for
urination in mammals simply scales with linear size, and consider the
scenarios where the driving force is gravity or elasticity, and where the
response is dominated by inertia or viscosity. We ask how the time required
for urination depends on the length scale, and find that for the time to be
independent of body size, the dominant driving force must be elasticity, and
the dominant response viscosity. Our note demonstrates that dimensional
analysis can indeed readily give insights into complex physical and biological processes.

\end{abstract}
\date{March 1, 2014}
\maketitle
\preprint{HEP/123-qed}
\affiliation{Liquid Crystal Institute, Kent State University}
\affiliation{Liquid Crystal Institute, Kent State University}
\affiliation{Liquid Crystal Institute, Kent State University}
\affiliation{Dept. of Mathematical Sciences, Kent State University}
\volumeyear{year}
\volumenumber{number}
\issuenumber{number}
\eid{identifier}
\received[Received text]{date}

\revised[Revised text]{date}

\accepted[Accepted text]{date}

\published[Published text]{date}

\startpage{1}
\endpage{ }

\section{Introduction}

In a recent publication [1], David Hu and coworkers report the discovery of
the "Law of Urination", according to which animals empty their bladders in the
nearly constant time interval of $21\pm13$ $s\,$regardless of size. This is
indeed a remarkable result, given that body masses of the animals in question,
from mice to elephants, range over some five orders of magnitute. The authors
go on to say that this feat is made possible by the increasing urethral length
of large animals, which 'amplifies gravitational force and flow rate'.
Although the dispersion about the mean time is certainly significant, and the
range of linear sizes of the animals considered, taken as cube root of the
volume, is about one order of magnitude, the Law of Urination is still
striking. In order to gain some insights into this phenomenon, we turn to
dimensional analysis.

Dimensional analysis is an exceedingly efficient and powerful tool of physics,
enabling insights into complex problems with relatively small computational
effort. In the hands of expert practitioners, it can be said to rise to the
level of art. Two inspirational examples of its use are the estimate of the
yield of the Trinity nuclear test by G.I.\ Taylor [2] and the estimate of the
height of mountains of earth by V. Weisskopf [3] and Goldreich \textit{et al} [4].

Dimensional analysis is based on the notion that the laws of physics\ must
have the same form in any system of units.\ This implies that
the\ relationship between the physical variables describing the phenomenon
under consideration can be expressed in terms of quantities without units. The
art of dimensional analysis lies in determining the relevant physical
variables, including only, but all, of the essential ones. In general, a
number of independent dimensionless groups can be formed; a formal procedure
for obtaining these is provided by the Buckingham $\Pi$ theorem [5].\ Here we
adopt a minimalist approach, and select only sufficient physical variables to
form a single dimensionless group, and use this simple dimensional analysis to
examine the role of gravity and other factors in the time needed for mammals
to urinate.

\section{Analysis}

We assume, for the pupose of this analysis, that the relative dimensions of
the bladder and urethra do not change from animal to animal, but the entire
structure scales with the linear size of the animal in question. Specifially,
we want to determine the time needed to empty a compact bladder, assumed to be
spherical, via a straight tube, the urethra, with circular cross-section.\ A
simple illustration is shown in Fig. 1.

\begin{figure}[th]
\begin{center}
\includegraphics[width=8.2cm]{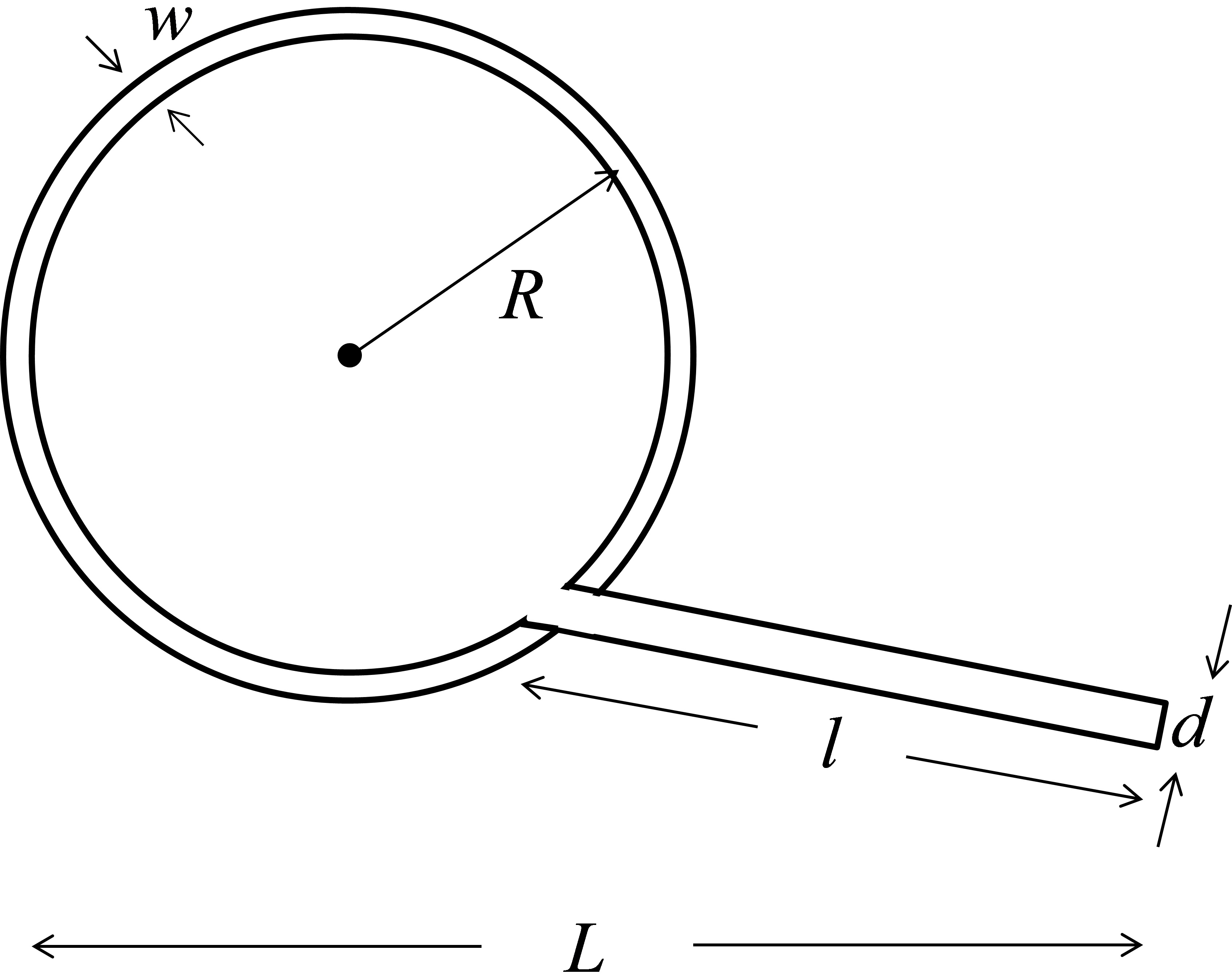}
\end{center}
\caption{Schematic of bladder and urethra.}%
\end{figure}

As pointed out in [1], some authors ascribe the force which is responsible for
expelling the urine to bladder pressure [6-8], while others propose a
combination of gravity and bladder pressure [9]. For a given driving force and
geometry, the flow velocity is determined primarily by fluid inertia and
viscosity. In our simple dimensional analysis, we therefore consider,
separately, the effects of the driving forces of gravity and muscle
contraction, and of the inertial and viscous response.

The essence of our dimensional analysis is to find how the relevant quantities
can be combined to form a dimensionless group, or, more specifially, how the
parameters describing the dominant driving force and response can be combined
to form a quantity with the units of the quantity of interest: the urination time.

\subsection{Gravity and Inertia}

Here we assume that the dominant driving force for the flow is the force of
gravity on the fluid, and the response is dominated by inertia. In this case,
the time $t_{u\text{ }}$to urinate, that is, to empty the bladder is a
function of the gravitational force per volume $\rho g$ acting on the fluid,
where $\rho$ is the mass density of urine and $g$ is the acceleration of
gravity, of inertia, as measured by the mass density $\rho$, and of the
various lengths illustrated in Fig.1. There $R$ is the radius of the bladder,
$w$ is the thickness of the detrusor muscle, $l$ is the length and $d$ is the
diameter of the urethra. Since all lengths in the problem can be written as
some dimensionless factor times a characteristic length $L$, we consider
$t_{u}$ to be a function of $\rho g$, $\rho$ and $L$. If we express units of
all relevant quantities in the problem in terms of the fundamental dimensional
quantities of mass $M$, time $T$, length $\mathcal{L}$, we have%
\begin{equation}
\lbrack\rho g]=\frac{M}{\mathcal{L}^{2}T^{2}},
\end{equation}
where the square brackets $[]$ indicate 'units (dimensions) of ',%
\begin{equation}
\lbrack\rho]=\frac{M}{\mathcal{L}^{3}},
\end{equation}
and $[L]=\mathcal{L}.\ $To form an expression with the unit of time, we must
determine the exponents $\alpha,\beta$ and $\gamma$ which satisfy%
\begin{equation}
T=[(\rho g)^{\alpha}\rho^{\beta}L^{\gamma}]=\frac{M^{\alpha}}{\mathcal{L}%
^{2a}T^{2\alpha}}\frac{M^{\beta}}{\mathcal{L}^{3\beta}}\mathcal{L}^{\gamma}.
\end{equation}
Equating exponents on both sides, we get at once $\alpha=-\frac{1}{2}%
,\beta=\frac{1}{2}$ and $\gamma=1/2$, so
\begin{equation}
t_{u}=\sqrt{\frac{L}{g}}\times C_{1},
\end{equation}
where $C_{1}\,$is a dimensionless function of the ratios of lengths in Fig. 1.
Since $C_{1}$ does not change if all lengths are scaled, it follows that, if
the dominant contributions were gravity and inertia, the time needed to
urinate would be independent of the density of urine but proportional to the
square root of the characteristic length of the animal (or proportional to the
mass to the one-sixth power, in accordance with [1] for large animals); large
animals would take somewhat longer to urinate than small ones, and the
required time would not be size independent.

\subsection{Gravity and Viscosity}

Here we assume that the dominant driving force is\ again gravity, but
the\ response is dominated by viscosity. The units of dynamic viscosity $\mu$
are, in SI units, $Pa-s$, or in terms of fundamental dimensional quantities%
\begin{equation}
\lbrack\mu]=\frac{M}{\mathcal{L}T}.
\end{equation}
Proceeding as before, we have%
\begin{equation}
T=[(\rho g)^{\alpha}\mu^{\beta}L^{\gamma}]=\frac{M^{\alpha}}{\mathcal{L}%
^{2a}T^{2\alpha}}\frac{M^{\beta}}{\mathcal{L}^{\beta}T^{\beta}}\mathcal{L}%
^{\gamma},
\end{equation}
ans we find that $\alpha=-1$, $\beta=1$ and $\gamma=-1$, so%
\begin{equation}
t_{u}=\frac{\mu}{\rho gL}\times C_{2},
\end{equation}
where $C_{2}$\thinspace again is a scale independent dimensionless function of
length ratios. Here, the urination time $t_{u}$ is inversely proportional to
the characteristic length of the animal; so large animals would take much less
time to urinate than small ones.

\subsection{Elasticity and Inertia}

We next assume that the dominant driving force is detrusor muscle tension,
which is characterized by the stress $Y$, and that the\ response is inertia
dominated. The dimensions of stress are%
\begin{equation}
\lbrack Y]=\frac{M}{\mathcal{L}T^{2}}.
\end{equation}
Proceeding as before,
\begin{equation}
T=[Y^{\alpha}\rho^{\beta}L^{\gamma}]=\frac{M^{\alpha}}{\mathcal{L}^{\alpha
}T^{2\alpha}}\frac{M^{\beta}}{\mathcal{L}^{3\beta}}\mathcal{L}^{\gamma},
\end{equation}
and we find that $\alpha=-\frac{1}{2}$, $\beta=\frac{1}{2}$ and $\ \gamma=1$,
so%
\begin{equation}
t_{u}=L\sqrt{\frac{\rho}{Y}}\times C_{3}.
\end{equation}
The time is thus proportional to the characteristic length; again at variance
with observations. In this scenario, large animals would take much longer to
urinate than small ones.

\subsection{Elasticity and Viscosity}

Lastly, we consider the case when the driving force is detrusor muscle
tension, and the response is viscosity dominated. Proceeding again as before,
\begin{equation}
T=[Y^{\alpha}\mu^{\beta}L^{\gamma}]=\frac{M^{\alpha}}{\mathcal{L}^{\alpha
}T^{2\alpha}}\frac{M^{\beta}}{\mathcal{L}^{\beta}T^{\beta}}\mathcal{L}%
^{\gamma},
\end{equation}
and we find that $\alpha=-1$, $\beta=1$ and $\gamma=0$, so%
\begin{equation}
t_{u}=\frac{\mu}{Y}\times C_{4},
\end{equation}
that is, the time is independent of characteristic length of the animal, in
accordance with observations.\ According to this result, large and small
animals take approximately the same time to urinate, regardless of size.

\section{Discussion and Conclusions}

In our simple model, we have assumed that in the animals considered, the
geometry of bladder and urethra is unchanged, and all lengths scale with the
characteristic length of the animal. We have considered separately the cases
when the dominant driving force for urination is gravity and muscle
contraction, and the dominant response is inertial and viscous.\ In three of
the four cases considered, the time required for urination depends on the
animal size; only in the case of muscle contraction and a viscous response do
we find that the time for urination is\ size independent. This is in agreement
with experimental observations. A rough estimate of the urination time, given
in the Appendix, is in reasonable agreement with experimental observations.
Our simple model therefore suggests that urination depends primarily on muscle
contraction and viscosity; gravity and inertia play a less important role.
This conclusion is in disaccord with the argument of Hu et al [1], who
stresses the importance of gravity in large animals in explaining experimental
observations. We note that astronauts apparently urinate without difficulty
even in the absence of gravity [8]; and babies pee in the upward direction as
well as down. If indeed the dominant factors are muscle contraction and
viscosity, as we argue here, then, according to our simple analysis in the
Appendix, the time is proportional to the ratio of urethral length to
diameter.\ This suggests that the urination time of females should be
significantly less than (about $1/5$ of) that of males of the same size. We
are currently seeking information and data to test this prediction.\ The
factor of $5$ is nearly within the $8-34s$ interval cited in [1].

Our estimate of times in the Appendix does not take a key aspect of urination
into account: namely, that the uretrha is compliant, and its diameter is a
nonlinear function of pressure [9]. This is a fascinating aspect, which
suggests non-steady flow, and energy transfer to the surrounding
tissue.\ Since these aspects have not been taken into account, the expressions
for urination times in the Appendix cannot be accurate, but they do indicate
how the geometry enters the dimensionless factor. We note that, even in the
case of compliant\ urethra, our dimensionless analysis is valid, since the
modulus of tissue is comparable to muscle stress, and thus there are no new
dimensional quantities entering the problem.

Finally, we note that for small animals, our analysis does not hold, since
other factors, such as surface tension, come into play.

In summary, we have shown here that simple dimensional analysis, with little
affort can give interesting and useful insights into complex phenomena, in
this case, the Law of Urination.

\section{Appendix}

In this section, we estimate the urination times using physical, rather than
dimensional arguments. Since this approach gives the dimensionless
multiplicative constant explicitly, it enables rough estimation of the
required times. We note that our simple approach ignores the compliance of the urethra.

\subsection{Gravity and Inertia}

On equating gravitational potential energy with kinetic energy in the urethra,
we get
\begin{equation}
\rho gR=\frac{1}{2}\rho v_{ua}^{2},
\end{equation}
where $v_{ua}$\ is the average velocity of the fluid in the urethra. The time
requried to empty the bladder is of the order
\begin{equation}
t_{u}=\frac{\frac{4}{3}\pi R^{3}}{\pi(\frac{d}{2})^{2}v_{ua}}=\sqrt{\frac
{R}{g}}\times\frac{16}{3\sqrt{2}}(\frac{R}{d})^{2},
\end{equation}
which is proprtional to the square root of the characteristic length. This is
at variance with the observation that the time is length independent.
Interestingly, $t_{u}$ is independent of the length of the urethra; it only
depends on its diameter.

Evaluating this using length estimates for humans gives $t_{u}\simeq118s.$
($72s$ for a cat, and $195s$ for an elephant.)

\subsection{Gravity and Viscosity}

In the Stokes limit for pipe flow with no slip at the boundaries, the local
velocity $v_{u}$ in the urethra satisifies
\begin{equation}
\mu\nabla^{2}v_{u}(r)=-\nabla P=-P^{\prime},
\end{equation}
and, assuming cylindrical symmetry,
\begin{equation}
\mu\frac{1}{r}\frac{\partial}{\partial r}(r\frac{\partial v}{\partial
r})=-P^{\prime},
\end{equation}
where the pressure gradient $P^{\prime}$ is a constant. The parabolic velocity
profile%
\begin{equation}
v_{u}=v_{o}(1-(\frac{r}{R_{o}})^{2})
\end{equation}
satisfies the equation, and substitution gives%
\begin{equation}
v_{o}=\frac{R_{o}^{2}P^{\prime}}{4\mu}.
\end{equation}
The volume current density $J$\ is just the velocity $v$. The flux $f$
(volume/time)\ is
\begin{equation}
f=\int JdA=\int_{0}^{R_{o}}v2\pi rdr=2\pi v_{o}\int_{0}^{R_{o}}(1-(\frac
{r}{R_{o}})^{2})rdr=\frac{\pi}{2}v_{o}R_{o}^{2},
\end{equation}
and substituting for $v_{o}$, we get for the flux%
\begin{equation}
f=\frac{\pi}{8}\frac{R_{o}^{4}}{\mu}P^{\prime}.
\end{equation}
If $R_{o}=d/2\ $and $P^{\prime}=P/l$, we get%
\begin{equation}
f=\frac{\pi}{128}\frac{d^{4}}{l}\frac{P}{\mu}.
\end{equation}
Estimating the pressure as%
\begin{equation}
P=\rho gR,
\end{equation}
then the required time is%
\begin{equation}
t_{u}=\frac{\frac{4}{3}\pi R^{3}}{f}=\frac{\frac{4}{3}\pi R^{3}l\mu}{\frac
{\pi}{128}d^{4}\rho gR}=\frac{\mu}{\rho gR}\times\frac{512}{3}(\frac{R}%
{d})^{3}\frac{l}{d}.
\end{equation}
Evaluating this using estimates for humans gives $t_{u}\simeq92.8s$.\ ($252s$
for a cat, and $34s$ for an elephant.)

\subsection{Elasticity and Inertia}

We can estimate the time using energy conservation in this case. The pressure
due to muscle contraction is%
\begin{equation}
P=\frac{2Yw}{R},
\end{equation}
and energy conservation gives, for the average velocity in the urethra,
\begin{equation}
v_{ua}=\sqrt{\frac{2P}{\rho}}=\sqrt{\frac{4Yw}{R\rho}},
\end{equation}
and the time to urinate is%
\begin{equation}
t_{u}=\frac{\frac{4}{3}\pi R^{3}}{\pi(\frac{d}{2})^{2}v_{ua}}=\frac{\frac
{4}{3}\pi R^{3}}{\pi(\frac{d}{2})^{2}}\sqrt{\frac{R\rho}{4Yw}}=R\sqrt
{\frac{\rho}{Y}}\times\frac{8}{3}(\frac{R}{d})^{2}\sqrt{\frac{R}{w}},
\end{equation}
and evaluating this using length estimates for humans gives $t_{u}=35.1s$.
($13s$ for a cat, and $95s\ $for an elephant.)

\subsection{Elasticity and\ Viscosity}

The flux in the urethra is, again
\begin{equation}
f=\frac{\pi}{128}\frac{d^{4}}{l}\frac{P}{\mu},
\end{equation}
and the pressure in the bladder is, again,
\begin{equation}
P=\frac{2Yw}{R}.
\end{equation}
The time to urinate therefore is
\begin{equation}
t_{u}=\frac{\frac{4}{3}\pi R^{3}}{f}=\frac{128\frac{4}{3}\pi R^{4}l\mu}{2\pi
d^{4}Yw}=\frac{\mu}{Y}\times\frac{256}{3}(\frac{R}{d})^{4}\frac{l}{w}.
\end{equation}
Evaluating this using length estimates for humans gives $t_{u}=8.2s$,
regardless of size, which is the right order of magnitude. \ This results has
a remarkably strong dependence - fourth power - of the urination time on
length ratios. Given the uniformity of experimental times, some other
mechanism (such as the dependence of the effective diameter of the urethra on
flow velocity [10]) most likely also contributes to the flow regulation in
biological systems.

\subsection{Physical Parameters}

The estimates below are for humans, and they are as follows:

$g=9.81m/s^{2}$

$R=6\times10^{-2}m$

$d=3\times10^{-3}m$

$l$ $=0.12m$ \ \ \ (for a male, $l=20cm$, for a female, $l=4cm$, $12cm$ is the average.)

$w=2\times10^{-3}m$

$\rho=10^{3}kg/m^{3}$

$Y=1\times10^{5}Pa$

$\mu=1\times10^{-3}Pa\ s$

$m=80kg$ (human)

$m=4kg$ (cat)

$m=4000kg$ (elephant)

We estimate characteristic lengths on basis of body mass. \ The ratio of the
characteristic lenghts of an elephant to that of a human is $\sqrt[3]%
{4000/80}=3.684$, and the ratio of a human to a cat is $\sqrt[3]{80/4}=2.71$.
By this measure, range of length scales covered is $3.684\times2.71\simeq10$.

\section{References}

1. P. J. Yang, J.C. Pham, J. Choo and D.L. Hu, "Law of Urination: all mammals
empty their bladders over the same duration",\ arXiv:1310.3737v2.

2. G. I. Taylor, "The dynamics of the combustion products behind plane and
spherical detonation fronts in explosives.", \textit{P. Roy. Soc. A-Math.
Phy.,}\textbf{\ 200}, 235-247 (1950).

3. V. Weisskopf,\ "\ Of Atoms, Mountains, and Stars: A Study in Qualitative
Physics", \textit{Science}, \textbf{187, }605-612 (1975)

4. P. Goldreich, S. Mahajan and S. Phinney, "Order-of-Magnitude Physics:
Understanding the World with\ Dimensional Analysis, Educated Guesswork and
White Lies", http://www.inference.phy.cam. ac.us/sanjoy/oom/book-letter.pdf

5. A. A. Sonin, 'The physical basis of dimensional analysis', 2001. This
manuscript is available from http://me.mit.edu/people/sonin/html.

6.\ S.G. Rao, J.S. Walter, A. Jamnia, J.S. Wheeler and M.S. Damaser,
"Predicting urethral area from video-urodynamics in women with voiding
dysfunction.", \textit{Neurourol. Urodyn.,} \textbf{22}, 277-283 (2003).

7.\ J. Walter, J. Wheeler, C. Morgan and M. Plishka, "Urodynamic evaluation of
urethral opening area in females with stress incontinence.", \textit{Int.
Urogynecol. J.,} \textbf{4}, 335-341 (1993).

8. H. Hollins, "Forgotten Hardware: How to Urinate in a Space Suit",
\ \textit{Adv. Physiol. Educ.} \textbf{37}, 123-128 (2013).

9. O. Barnea and G. Gillon, "Model-based estimation of male urethral
resistance and elasticity using pressure-flow data.", \textit{Comput. Biol.
Med.}, \textbf{31}, 27-40 (2001).

10. J.A. Martin, and S.S. Hillman, "The physical movement of urine from the
kidneys to the urinary bladder and bladder compliance in two anurans.",
\textit{Physiol. Biochem. Zool.,} \textbf{82}, 163-169 (2009).

\end{document}